\documentclass[prb,aps,showpacs,floatfix,floats,nobalancelastpage,twocolumn]{revtex4}
\usepackage{amsmath, euscript}
\usepackage{amsfonts}
\usepackage{amssymb}
\usepackage{graphicx}
\usepackage{color}
\usepackage{dcolumn}
\usepackage{bm}
\usepackage{subfigure}
\usepackage{ulem}

\def\ra{\rangle}
\def\la{\langle}
\def\Hc{{\rm H.c.}}
\def\bk{{\bf k}}
\def\br{{\bf r}}

\def\cB{{\cal B}}
\def\cL{{\cal L}}
\def\cM{{\cal M}}
\def\Gutz{{\rm Gutzw}}
\def\EBL{{\rm EBL}}

\begin{document}

\title{Failure of Gutzwiller-type wave function to capture gauge fluctuations: Case study in the Exciton Bose Liquid context}

\author{Tiamhock Tay}
\author{Olexei I. Motrunich}

\affiliation{Department of Physics, California Institute of Technology, Pasadena, CA 91125}

\date{\today}


\begin{abstract}
Slave particle approaches are widely used in studies of exotic quantum phases. A complete description beyond mean field also contains dynamical gauge fields, while a simplified procedure considers Gutzwiller-projected trial states. We apply this in the context of bosonic models with ring exchanges realizing so-called Exciton Bose Liquid (EBL) phase and compare a Gutzwiller wave function against an accurate EBL wave function. We solve the parton-gauge theory and show that dynamical fluctuations of the spatial gauge fields are necessary for obtaining qualitatively accurate EBL description. On the contrary, just the Gutzwiller projection leads to a state with subtle differences in the long-wavelength properties, thus suggesting that Gutzwiller wave functions may generally fail to capture long-wavelength physics.
\end{abstract}
\maketitle

\section{Introduction}

Over the past decade, the study of spin liquids and non-Fermi liquids have been an active theme in condensed matter physics.\cite{LeeNagaosaWen2006, Balents2010}  A common approach used in many of these studies involves the notion of fractionalization where the original particles of a microscopic model are substituted with slaves particles coupled to a gauge field.\cite{Wen2002, LeeNagaosaWen2006}  It is often thought that Gutzwiller wave functions, constructed by performing projection into the physical Hilbert space, are able to capture the correct physics.  However, it is suspected that such wave functions may be not sufficient to capture the long wavelength properties in important cases with gapless gauge fields, e.g.\ for U(1) spin liquids, \cite{LeeNagaosaWen2006, Holstein, Reizer, LeeNagaosa, Polchinski94, Altshuler94, YBKim94, SSLee2009, Metlitski2010, Mross2010, Rantner2002, Hermele2004} as the Gutzwiller construction does not include spatial gauge fluctuations.\cite{Motrunich2005, HermeleRan2008}
In a recent study of a hard-core boson model with pure ring exchange interactions which we proposed as a candidate model for realizing an Exciton Bose Liquid (EBL) phase,\cite{Paramekanti2002, Xu2005, Xu2007} we noticed that the EBL can be viewed as a special solvable example of a gapless parton-gauge system.\cite{Tay2010a, Tay2010b}  In this work, we shall take up this critical issue that Gutzwiller wave functions might not capture the spatial gauge fluctuations by explicit demonstrations in the EBL context.

To set the stage for our discussion, we begin with a schematic hard-core boson model with ring exchange interactions which serves the dual-purpose of introducing the EBL theory as well as motivating the wave functions used in this work. The Hamiltonian defined on the square lattice is
\begin{eqnarray}
  H_{\rm ring} &=& -\sum_{\br,m,n} [K_{mn} P_{mn}(\br) + \Hc],  \label{eq:ring_hamiltonian}\\
  P_{mn}(\br) &=& b^\dagger_\br ~b_{\br+m\hat{x}} ~b^\dagger_{\br+m\hat{x}+n\hat{y}} ~b_{\br+n\hat{y}},
\end{eqnarray}
where $P_{mn}(\br)$ are extended ring exchanges on $m\times n$ plackets and $K_{mn}$ are amplitudes for these exchanges.  The Hamiltonian conserves boson number on each row and column, and throughout our Hilbert space is the sector with equal number of bosons on each row and column.  We will assume $K_{mn} \geq 0$.  Our recent study of a model with $K_{11}$ and $K_{12}=K_{21}$ found regimes of the EBL phase,\cite{Tay2010a,Tay2010b} while here we are not concerned with a detailed realization but rather qualitative aspects, assuming the model Eq.~(\ref{eq:ring_hamiltonian}) is deep in the EBL phase.

In the following, we consider a slave-particle approach applied to this problem. Writing each boson operator as a product of two parton operators
\begin{equation}
b^\dagger_\br = b^\dagger_{\br1} ~b^\dagger_{\br2},
\end{equation}
we can recover the physical Hilbert space by imposing the constraint $n(\br)=n_1(\br)=n_2(\br)$. We then consider states where the $b_1$ partons hop only in the $\hat{x}$ direction while $b_2$ partons hop only in the $\hat{y}$ direction\cite{Motrunich2007} (so a single microscopic boson $b$ indeed cannot hop by itself), and further justify this by noting that the mean-field expectation value of each ring term in Eq.~(\ref{eq:ring_hamiltonian}) acquires a large, negative energy
\begin{eqnarray}
  \left\la -K_{mn} P_{mn}(\br) \right\ra_{\rm m.f.} &=& -K_{mn} \left\vert G_1(m\hat{x}) \right\vert^2 \left\vert G_2(n\hat{y}) \right\vert^2, \label{eq:ring_energy}\\
  G_\mu(m\hat{\mu}) &\equiv& \left\la b^\dagger_{\br\mu} b_{\br+m\hat{\mu},\mu} \right\ra_{\rm m.f.}. \label{eq:parton_propagator}
\end{eqnarray}

Beyond the mean field slave particle treatment, we introduce fluctuations into the theory by coupling the two parton species to a gauge field  ${\bm a}$ residing on the links of the lattice, with opposite gauge charges for the respective species.  The parton-gauge system is qualitatively captured by the following U(1) lattice gauge theory\cite{Motrunich2007}
\begin{eqnarray}
  H_{\rm U(1)} &=& -t \sum_{\br,\mu} \left[e^{iq_\mu a_{\br\mu}}b^\dagger_{\br\mu} b_{\br+\hat{\mu},\mu} + \Hc\right]\label{eq:parton_hamiltonian}\\
  &+& h\sum_{\br,\mu} e^2_{\br\mu} - K\sum_\br \cos(\nabla\times {\bm a})_\br,\label{eq:gauge_hamiltonian}\\
  (\nabla\cdot {\bf e})_\br &=& \sum_{\mu} q_\mu b^\dagger_{\br\mu}b_{\br\mu} ~,\label{eq:Gauss_law}
\end{eqnarray}
where $q_\mu=\pm 1$ (${\mu=1,2}$ or ${x,y}$) are the gauge charges for the partons moving respectively along $\hat{x}$ and $\hat{y}$ directions. Equations~(\ref{eq:parton_hamiltonian}) and (\ref{eq:gauge_hamiltonian}) are the respective Hamiltonians for the partons and the gauge fields, while Gauss' law in Eq.~(\ref{eq:Gauss_law}) imposes a constraint on the physical states.  The lattice curl and divergence used in Eqs.~(\ref{eq:gauge_hamiltonian}) and (\ref{eq:Gauss_law}) are defined by
\begin{eqnarray}
  (\nabla\times{\bm a})_\br &=& a_{\br+\hat{x},y} - a_{\br y} - a_{\br+\hat{y},x} + a_{\br x} ~,\\
  (\nabla\cdot{\bf e})_\br &=& e_{\br x} + e_{\br y} - e_{\br-\hat{x}, x} - e_{\br-\hat{y},y} ~.
\end{eqnarray}

In the above U(1) gauge theory, the integer-valued ``electric'' field $e_{\br\mu}$ is canonically conjugate to the compact gauge field $a_{\br\mu}$ on the same lattice link. Dynamical fluctuation of these fields arises from the competing terms in the gauge field Hamiltonian. In the limit $h\gg K,t$, the electric field vanishes and the Gauss' law reduces to $n_1(\br)=n_2(\br)$, which projects back into the physical boson Hilbert space. In this limit, it is possible to eliminate the gauge field perturbatively and obtain a Hamiltonian for hard-core bosons on the square lattice with ring exchange terms of the type in Eq~(\ref{eq:ring_hamiltonian}), thus establishing formal connection between $H_{\rm ring}$ and $H_{\rm U(1)}$.\cite{Motrunich2007}

As we will argue below, the EBL phase in $H_{\rm ring}$ corresponds to a ``deconfined" phase of $H_{\rm U(1)}$, where we can ignore the compactness of the gauge field, and treat the spatial gauge fluctuations fully. This is possible in the present case due to the powerful bosonization technique made applicable by the one-dimensional character of the partons and some ``dimensional reduction'' occuring in the system.\cite{Paramekanti2002, Xu2005, Xu2007, Nussinov2005, Batista2005, Nussinov2006} On the other hand, a different route beyond mean field often used in the literature is to apply Gutzwiller projection, mostly popular because of its numerical tractability\cite{Ceperley77, Gros89} (while gauge theories are often intractable). As one can anticipate, this state does not know about the spatial gauge field fluctuations and fails to reproduce the long-distance properties of the EBL phase.

The paper is organized as follows.
In Sec.~\ref{sec:theories}, we start from a Lagrangian formulation and show how the gauge theory leads to the EBL field theory, while neglecting the spatial gauge field fluctuations leads to a decidedly different low energy effective theory.  In Sec.~\ref{sec:trial_wavefunction}, we construct the wave functions used in this paper, and derive results for density structure factor and box correlator in the harmonic approximation for the wave functions.  In Sec.~\ref{sec:numerics}, we present our accurate Variational Monte Carlo (VMC) calculations for hard-core bosons and show that the Gutzwiller wave function indeed realizes a quantum state that is distinct from the EBL. In the conclusion, we discuss our study more broadly.

\section{Effective actions for the EBL and Gutzwiller theories}\label{sec:theories}

For the remainder of the paper, we assume a stable ``deconfined'' phase of $H_{\rm U(1)}$ where we can ignore compactness of the parton phase variables and compactness of the gauge field (stability is discussed in Appendices A,B of Ref.~\onlinecite{Motrunich2007}, borrowing from stability analyses of the EBL in Refs.~\onlinecite{Paramekanti2002, Xu2005, Xu2007}). To study the qualitative effects of spatial gauge fluctuations, we consider the following parton-gauge Lagrangian which provides a transparent starting point for our analysis
\begin{eqnarray}
  \cL &=& \frac{v}{2\pi}\left[ g^{-1}(\partial_x\theta_1)^2 + g(\partial_x\phi_1-a_x)^2\right] + \frac{i}{\pi}(\partial_x\theta_1)(\partial_\tau\phi_1)\nonumber\\
  &+& \frac{v}{2\pi}\left[ g^{-1}(\partial_y\theta_2)^2 + g(\partial_y\phi_2+a_y)^2\right] + \frac{i}{\pi}(\partial_y\theta_2)(\partial_\tau\phi_2) \nonumber\\
  &+& \frac{\kappa}{2} (\partial_x a_y - \partial_y a_x)^2,\label{eq:parton_gauge}
\end{eqnarray}
where the coarse-grained fields $\phi_\mu$ and $\theta_\mu$ provide a hydrodynamic fluid description of partons moving in the respective direction $\hat{\mu}$, and minimally coupled to the gauge field ${\bm a}$.  The velocity $v$ and dimensionless parameter $g$ are convenient parametrization from the bosonization literature.\cite{Haldane1981,Fisher1989,Fisher1998,Giamarchi} In this formulation, $\phi_\mu$ gives the phase of a parton while the dual variable $\theta_\mu$ is related to the parton density fluctuation through $\delta n_\mu = \pi^{-1}\partial_\mu \theta_\mu$.  We also assume a sizable ``stiffness'' $\kappa$ for the gauge field (e.g., set by the energetics of the boson ring exchanges).  Instead of introducing the temporal gauge field, we impose the following constraint at each lattice site
\begin{equation}
  \partial_x\theta_1 = \partial_y\theta_2, \label{eq:constraint}
\end{equation}
which allows to recover the physical Hilbert space by binding two partons to give the original boson. The constraint is then solved by introducing a field $\vartheta$ satisfying
\begin{equation}
\theta_1 = \partial_y\vartheta, ~~~\theta_2 = \partial_x\vartheta, \label{eq:vartheta}
\end{equation}
where, as the analysis below shows, $\vartheta$ can be identified as the coarse-grained field dual to the boson phase $\phi$ in the ``bosonization'' of the two-dimensional ring exchange model in Ref~\onlinecite{Paramekanti2002}. 

We first integrate out the fields $\phi_\mu$ and obtain
\begin{eqnarray}
  \cL_{\rm eff} &=& \frac{v}{\pi g} (\partial_x\partial_y\vartheta)^2 + \frac{1}{2\pi vg}[(\partial_\tau\partial_x\vartheta)^2 + (\partial_\tau\partial_y\vartheta)^2] \nonumber\\
  &+& \frac{i}{\pi}(\partial_\tau\vartheta) (\partial_x a_y - \partial_y a_x) + \frac{\kappa}{2}(\partial_x a_y - \partial_y a_x)^2 .~~~~\label{eq:common_action}
\end{eqnarray}
After further integrating out the gauge field ${\bm a}$ and then dropping a less relevant term $(\partial_\tau {\bm \nabla} \vartheta)^2$, we arrive at the following realization of the EBL theory
\begin{equation}
  \cL_\EBL = \frac{1}{2\pi^2\kappa}(\partial_\tau\vartheta)^2 + \frac{v}{\pi g}(\partial_x\partial_y\vartheta)^2,\label{eq:EBL_action}
\end{equation}
where the more general EBL theory is defined by the action\cite{Paramekanti2002}
\begin{eqnarray}
  {\EuScript S}_\EBL[\vartheta] = \frac{1}{2} \sum_{\bk,\omega} \cM_\EBL(\bk,\omega) ~\vert\vartheta(\bk,\omega)\vert^2,\label{eq:EBL_theory}\\
  \cM_\EBL({\bf 0}, \omega) \sim \omega^2, ~~~\cM_\EBL(\bk,0) \sim \vert k_x k_y \vert^2, \label{eq:EBL_condition}
\end{eqnarray}
for small $k_x$, $k_y$.  [Strictly speaking, going from Eq.~(\ref{eq:common_action}) to Eq.~(\ref{eq:EBL_action}), we need to keep $\cM_\EBL(\bk, \omega)$ accurately on the full lines $\bk = (0, k_y)$ and $(k_x, 0)$, i.e., we should not drop the naively less relevant term $(\partial_\tau {\bm \nabla} \vartheta)^2$.  However, here we focus on long-wavelength effects originating near $\bk = (0, 0)$ and work in a schematic continuum notation, while an accurate lattice variant can be found in Appendix B of Ref.~\onlinecite{Tay2010b}]. The energy dispersion can be obtained from Eq.~(\ref{eq:EBL_condition}) and has the form $E_\bk\sim |k_xk_y|$. This is responsible for interesting properties of the EBL phase\cite{Paramekanti2002} such as specific heat $C\sim T\log(1/T)$, which makes it qualitatively different from sliding or cross-sliding Luttinger liquid phases.\cite{Emery2000, Vishwanath2001, Ranjan2001long} 
[Generally, the vanishing of $E_\bk$ along the lines $(0, k_y)$ and $(k_x, 0)$  can be shown to be a consequence of the conservation of boson number in each row and column of the lattice ring model, and is satisfied in this parton-gauge approach by construction.]

Let us now see what happens if we do not have dynamical gauge fields. To obtain the resulting Lagrangian, we drop the gauge field from Eq.~(\ref{eq:common_action}):
\begin{equation}
  \cL_\Gutz = \frac{v}{\pi g} (\partial_x\partial_y\vartheta)^2 + \frac{1}{2\pi vg}[(\partial_\tau\partial_x\vartheta)^2 + (\partial_\tau\partial_y\vartheta)^2]. \label{eq:Gutz_Lagrangian}
\end{equation}
We will view this as a schematic model of what happens under Gutzwiller projection, hence the label ``Gutzw''.  The corresponding action is
\begin{eqnarray}
  {\EuScript S}_\Gutz[\vartheta] &=& \frac{1}{2} \sum_{\bk,\omega} \cM_\Gutz(\bk,\omega) ~\vert\vartheta(\bk,\omega)\vert^2, ~\label{eq:Gutz_theory}\\
  \cM_\Gutz(\bk, \omega) &=& \frac{2v}{\pi g} |k_x k_y|^2 + \frac{1}{\pi vg} \omega^2 \bk^2. \label{eq:Gutz_condition}
\end{eqnarray}
Here, the energy dispersion is $E_\bk \sim |k_xk_y|/\vert\bk\vert$ and the distinct behavior in the vicinity of $\bk={\rm 0}$ leads to low energy properties different from the corresponding EBL properties. For example, the specific heat vanishes linearly with temperature for the Gutzwiller action, i.e., does not have the logarithmic factor $\log(1/T)$ found for the EBL case.

The long wavelength properties of the EBL and Gutzwiller actions are also different. To give examples of other observable consequences, we calculate the density structure factor $D(\bk)$ and box correlator $\cB(x,y)$ defined below:
\begin{eqnarray}
  D(\bk) &\equiv& \left\la \vert n_\bk \vert^2 \right\ra, \\
  \cB(x,y) &\equiv& \left\la e^{i\left[\phi(0,0)-\phi(x,0)+\phi(x,y) -\phi(0,y)\right]}\right\ra,\label{eq:boxcorr_def}\\
  &=& e^{-\frac{1}{2}\int\frac{d^2\bk}{(2\pi)^2} \vert 1-e^{ik_xx}\vert^2\vert 1-e^{ik_yy}\vert^2\la \vert \phi_\bk \vert^2 \ra} ~,
\end{eqnarray}
with $\phi_\br$ denoting the boson phase variable, $b_\br^\dagger \sim e^{i\phi_\br}$. Here, $\left\la \vert n_\bk \vert^2 \right\ra$ can be evaluated for the Gaussian action using $\delta n_\br = \pi^{-1} \partial_x \partial_y \vartheta_\br$, and we obtain
\begin{eqnarray}
  D_\EBL(\bk) &=&  \frac{1}{2}\sqrt{\frac{g\kappa}{2\pi v}} ~|k_xk_y|, \label{S_EBL} \\
  D_\Gutz(\bk) &=& \frac{g}{2\sqrt{2}\pi} ~|k_xk_y| / |\bk|, \label{S_Gutzw}
\end{eqnarray}
for small $k_x$, $k_y$.  The singularity in the structure factor is distinct at $\bk={\bf 0}$ for the two actions.  Specifically, at fixed $k_y$, $D(k_x \to 0, k_y) = C(k_y) |k_x|$, with $C_\EBL(k_y) \sim |k_y|$ for small $k_y$, but $C_\Gutz(k_y) \sim {\rm const}$ for small $k_y$.

Since $n_\br$ and $\phi_\br$ are canonically conjugate to each other, they satisfy the following ground state minimum uncertainty relation
\begin{equation}
  \sqrt{\left\la \vert n_\bk \vert^2 \right\ra} \sqrt{\left\la \vert \phi_\bk \vert^2 \right\ra} = 1/2, \label{eq:boxcorr}
\end{equation}
which allows to obtain the box correlator Eq.~(\ref{eq:boxcorr_def}).
We will focus on the regime $|x| \gg |y|$, where we find power law decay $\sim |x|^{-\eta(y)}$ with $y$-dependent exponents.  To determine the exponents for all $y$, we in fact need to have details on the $(0, k_y)$ line all the way up to the Brillouin zone boundary [only the large $y$ limit is determined by focusing on the vicinity of $\bk = (0, 0)$].  For illustrations below, we simply take model $D_\EBL(\bk)$ and $D_\Gutz(\bk)$ by replacing $|k_y| \to 2|\sin(k_y/2)|$ in Eqs.~(\ref{S_EBL}) and (\ref{S_Gutzw}).
For the EBL case we find\cite{Paramekanti2002}
\begin{eqnarray}
  \cB_\EBL(x,y) &\sim& |x|^{-\eta_\EBL(y)}, \\
  \eta_\EBL(y) &=& \frac{1}{\pi^2}\sqrt{\frac{2\pi v}{g\kappa}}  \int_0^\pi \frac{\sin^2(k_y y/2)}{\sin(k_y/2)} ~dk_y \\
&\approx& \frac{1}{\pi^2}\sqrt{\frac{2\pi v}{g\kappa}} \log (y),
\end{eqnarray}
where the last line gives growth behavior for $|y| \gg 1$.
For the Gutzwiller box correlator we find
\begin{eqnarray}
  \cB_\Gutz(x,y) &\sim& A(y) ~|x|^{-\eta_\Gutz}, \\
  \eta_\Gutz &=& \sqrt{2}/g ~, \label{eta_Gutzw_action}
\end{eqnarray}
which is independent of $y$ in the present Gutzwiller model and generally remains finite for any $y$.

For finite $y$ and large $x$, $\cB(x,y)$ can be viewed as the propagator for an exciton of transverse size $y$. The qualitative difference in the box correlator for large transverse size shows that the two actions indeed lead to different long wavelength properties. Thus, whether or not one allows gauge fluctuation, does lead to effective low energy theories with distinct ground state properties.

We emphasize here that the stability of the EBL phase in the ring model given in Eq.~(\ref{eq:ring_hamiltonian}) is not the focus of this study. Instead, we take the parton-gauge action in Eq.~(\ref{eq:parton_gauge}) as our starting point and address the question whether excluding gauge fluctuation may lead to a qualitative difference.   Note that the Gutzwiller action is only a caricature of what happens under the Gutzwiller projection and one should use some effective parameter $g_{\rm eff}$ rather than bare $g$. In the next section, we will give a more accurate treatment by explicitly constructing a Gutzwiller wave function and comparing its properties with those of a model EBL wave function.

\section{Trial wave functions}\label{sec:trial_wavefunction}

In this section, we examine the formal properties of the Gutzwiller and EBL wave functions and highlight qualitative differences between them.  

\subsection{General Jastrow wave function and harmonic approximation}

We first derive expressions for the density structure factor and box correlator for a general Jastrow-type wave function with a two-body pseudo-potential\cite{Bijl1940,Jastrow1955}
\begin{equation}
  \Psi\left( \left\{ \br_i \right\} \right) \propto \exp\left[ -\frac{1}{2}\sum_{i,j}u(\br_i-\br_j)\right], \label{eq:Jastrow}
\end{equation}
where the indices $i,j$ run over the bosons. In the second-quantized notation on the lattice, the wave function can be equivalently expressed as
\begin{equation}
  \left\vert \Psi \right\ra \propto \sum_{\{n_\br\}} \exp{\left[-\frac{1}{2}\sum_{\br',\br''}u(\br'-\br'') n_{\br'} n_{\br''} \right]} \left\vert \{ n_\br\}\right\ra.~
\end{equation}
We will shortly see that both the EBL and Gutzwiller wave functions have such forms, and their pseudo-potentials $u(\br)$ will be given later.  If we disregard the discreteness of the boson number here, we obtain the following approximate density structure factor\cite{Reatto1967} for an arbitrary Gaussian wave function (viewed in $n_\br$ variable)
\begin{eqnarray}
  \left\la \vert n_\bk \vert^2 \right\ra &=& \frac{1}{2u_\bk},\\
  u_\bk &=& \sum_{\br} u(\br) e^{-i\bk\cdot\br}. 
\end{eqnarray}
The box correlator defined in Eq.~(\ref{eq:boxcorr_def}) can be calculated using
\begin{equation}
  \left\la \vert \phi_\bk \vert^2 \right\ra = \frac{1}{2}u_\bk, \label{eq:expectation_phi}
\end{equation}
which follows from the boson phase operator $\phi_\br$ being canonically conjugate to the boson number operator $n_\br$. Again, we have made use of the harmonic approximation, that is, we neglect the discreteness of $n_\br$, or equivalently the compactness of $\phi_\br$.

\subsection{EBL wave function}

For a model EBL wave function, we use the pseudo-potential from Refs.~\onlinecite{Motrunich2007, Tay2010b}, which can be motivated by a direct ``spin-wave'' treatment of the ring Hamiltonian in Eq.~(\ref{eq:ring_hamiltonian}),
\begin{equation}
  u_\EBL(\br) = \frac{1}{L^2}\sum_\bk \frac{W_\EBL ~e^{i\bk \cdot \br}}{4\vert \sin(k_x/2)\sin(k_y/2)\vert}.\label{eq:EBL_pseudo_potential}
\end{equation}
Note that we exclude lines $(k_x,0)$ and $(0,k_y)$ from the sum.  One can also turn this into a convergent integral by replacing $e^{ik_x x}$ by $e^{ik_x x}-1$ and $e^{ik_y y}$ by $e^{ik_y y}-1$; this does not change $\Psi_\EBL$ because of fixed particle number in each row and column (in our working Hilbert space appropriate for the ring models).  In principle, $W_\EBL$ can be a smooth function of $\bk$ but for simplicity here, we take it to be a constant.   We now use harmonic approximation and obtain the following density structure factor and box correlator
\begin{eqnarray}
  D_\EBL(\bk) &=& \frac{2}{W_\EBL} \left\vert \sin\left(\frac{k_x}{2}\right) \sin\left(\frac{k_y}{2}\right)\right\vert,\label{eq:EBL_sf}\\
  \cB_\EBL(x,y) &\sim& \vert x \vert^{-\eta_\EBL(y)},\label{eq:boxEBL}
\end{eqnarray}
where Eq.~(\ref{eq:boxEBL}) holds for large $x$ and fixed $y$ and $\eta_\EBL(y)$ is given by
\begin{eqnarray}
  \eta_\EBL(y) &=& \frac{W_\EBL}{\pi^2} \int_0^\pi \frac{\sin^2(k_y y/2)}{\sin(k_y/2)} ~dk_y\\
  &=& \frac{2W_\EBL}{\pi^2}\left[1+\frac{1}{3}+\frac{1}{5}+\cdots+\frac{1}{2y-1}\right]~~\\
  &\approx& \frac{W_\EBL}{\pi^2} \log(y), ~~~y\gg 1. \label{eq:log_approx}
\end{eqnarray}
Properties Eq.~(\ref{eq:EBL_sf}) with $D_\EBL(\bk) \sim |k_xk_y|$ and Eq.~(\ref{eq:boxEBL}) with $\eta_\EBL(y)$ growing logarithmically with $y$, are long wavelength properties of the EBL.\cite{Paramekanti2002}

\subsection{Gutzwiller wave function}

To obtain the Gutzwiller wave function, we use the following wave function for partons confined within a chain 
\begin{eqnarray}
  \Psi_{\rm chain}\left(\{x_i\}\right) &\propto& \exp\left[-\frac{1}{2}\sum_{i, j}u_{\rm 1d}(x_i-x_j)\right],\\
  u_{\rm 1d}(x) &=&\frac{1}{L}\sum_{k_x} \frac{W_{\rm 1d} ~e^{i k_x x}}{2\vert \sin(k_x/2)\vert}.
\end{eqnarray}
(We can again regularize the sum by replacing $e^{ik_x x}$ by $e^{ik_x x}-1$ since adding a constant to the pseudo-potential does not change the wave function for fixed particle number in the chain.)   This trial wave function has been known to capture the energetics as well as Luttinger liquid exponents of one-dimensional systems.\cite{Hellberg1991, Capello2005} We construct the Gutzwiller wave function as
\begin{equation}
  \Psi_\Gutz(\{\br_i\}) = \Psi_1(\{\br_i\})~\Psi_2(\{\br_i\}) ~,\label{eq:gutzwiller_wf}
\end{equation}
where $\Psi_\mu$ is the wave function for the $b_\mu$ partons confined to move within chains oriented in the $\hat{\mu}$ direction.  Note that Gutzwiller projection has been explicitly imposed in Eq.~(\ref{eq:gutzwiller_wf}) where both parton species are present at each boson location for any given set of $\{\br_i\}$. The Gutzwiller wave function indeed has a Jastrow form with the following pseudo-potential
\begin{equation}
  u_\Gutz(\br-\br') = \delta_{y,y'} ~u_{\rm 1d}(x-x') + \delta_{x,x'} ~u_{\rm 1d}(y-y').
\end{equation}
Again disregarding the discreteness of the boson numbers, we obtain the following density structure factor
\begin{equation}
  D_\Gutz(\bk) = \frac{1}{W_{\rm 1d}}\, \frac{\vert \sin(k_x/2) \sin(k_y/2)\vert}{\vert \sin(k_x/2)\vert + \vert\sin(k_y/2)\vert} ~.\label{eq:sfGutz}
\end{equation}
This has a different singularity at $\bk={\bf 0}$ compared to the density structure factor for the EBL wave function in Eq.~(\ref{eq:EBL_sf}).\cite{SingleBosonHopping}

The difference in the structure factors near $\bk={\bf 0}$ manifests itself in the fluctuation properties.    For a rectangular region $[0,x)\times[0,y)$, we define the following number fluctuation for the total number of bosons in the region
\begin{equation}
  \delta N(x,y) = \sum_{x'=0}^{x-1}\sum_{y'=0}^{y-1} \delta n(x',y') ~.
\end{equation}
The variance of the number fluctuation is readily calculated
\begin{equation*}
\la \delta N(x,y)^2 \ra = \frac{1}{L^2} \sum_\bk \left[\frac{\sin(k_x x/2) \sin(k_y y/2)}{\sin(k_x/2) \sin(k_y/2)}\right]^2 \left\la \vert n_\bk \vert^2 \right\ra ~.
\end{equation*}
In the limit $x \gg y \gg 1$, this has different asymptotic forms for the EBL and the Gutzwiller wave functions:
\begin{eqnarray}
  \left\la \delta N_\EBL(x,y)^2\right\ra &\approx& \frac{2}{\pi^2 W_\EBL} \, \log(x) \, \log(y),\label{eq:EBL_dN}\\
  \left\la \delta N_\Gutz(x,y)^2\right\ra & \approx& \frac{1}{\pi W_{1d}} \, y \, \log(x).
\end{eqnarray}
Equation~(\ref{eq:EBL_dN}) shows that such number fluctuation in the EBL wave function is strongly suppressed.
On the other hand, it scales linearly in the region width $y$ for the Gutzwiller case while increasing logarithmically with $x$.  This reminds of the additivity of variances of statistically independent random variables, and the Gutzwiller result appears to suggest that, in the absence of gauge fluctuations, the bosons in adjacent chains are weakly coupled compared to those in the EBL phase.  

We now turn to the box correlator and obtain
\begin{equation}
  \cB_\Gutz(x,y) \sim \vert x \vert^{-W_{\rm 1d}/\pi},\label{eq:boxGutz}
\end{equation}
for large $x$ and finite $y$, which is again qualitatively different from the EBL box correlator in that the exponent here does not grow with $y$.  Notice that the Gutzwiller result has in fact identical power law to the mean field box correlator [see Eq.~(\ref{eq:parton_propagator})]
\begin{eqnarray}
  \cB_{\rm m.f.}(x,y) &=& |G_1(x\hat{x})|^2~|G_2(y\hat{y})|^2,\\
  &\sim& |x|^{-W_{\rm 1d}/\pi}~|y|^{-W_{\rm 1d}/\pi}.
\end{eqnarray}
This therefore suggests that the Gutzwiller projection has not provided any improvement over the mean field slave particle treatment as far as long-distance properties are concerned.  [Note that the schematic treatment in Sec.~\ref{sec:theories} leading to results Eqs.~(\ref{S_Gutzw}) and (\ref{eta_Gutzw_action}) might suggest otherwise if we naively use $g = g_{\rm m.f.} = \pi/W_{1d}$ there; however, such treatment appears to over-emphasize the role of the constraint on the long-distance properties of the wave functions and we should allow some effective $g_{\rm eff}$ instead.  We believe the direct approach to the wave functions as in this section is more accurate and shows that there is no change in the power laws compared to the mean field.]

To conclude our harmonic approximation study of the EBL and the Gutzwiller wave functions in this section, we have shown that despite the ability of the Gutzwiller wave function to realize a quantum liquid, it does not give a fully qualitatively accurate representation of the EBL phase as defined by Eqs.~(\ref{eq:EBL_theory}) and (\ref{eq:EBL_condition}).  In the next section, we enforce hard-core boson condition at each lattice site in Variational Monte Carlo calculations and obtain numerically exact information for the corresponding wave functions, defined in sectors with fixed boson number in each row and column.

\section{Exact VMC Results}\label{sec:numerics}

In this section, we perform exact calculations for the hard-core bosons using the wave functions from Sec.~\ref{sec:trial_wavefunction}. We set up Variational Monte Carlo simulations which allow hard-core boson constraint to be imposed exactly.  We also require fixed boson number in each row and column.  Since we are only interested in wave functions which realize liquid phases, it is important to ensure that the variational parameter chosen for each trial wave function does not lead to an ordered phase.  For concreteness, we choose density with $\rho=1/2$ (i.e., $L/2$ bosons in each row and each column of $L\times L$ lattices), and select $W_\EBL=1.5$ for the EBL wave function and $W_{\rm 1d}=1.5$ for the Gutzwiller wave function so that both wave functions are deep inside the liquid regimes.\cite{Wcritical}  For the Monte Carlo random walks, we allow all possible $m\times n$ ring moves where bosons hop from occupied sites at $\br$ and $\br+m\hat{\bf x}+n\hat{\bf y}$ onto vacant sites at $\br+m\hat{\bf x}$ and $\br+n\hat{\bf y}$. These are the simplest moves that preserve the boson number in each row and column and also guarantee ergodicity in the Hilbert space of the problem.

\begin{figure}
  \centering
  \includegraphics[width=\columnwidth]{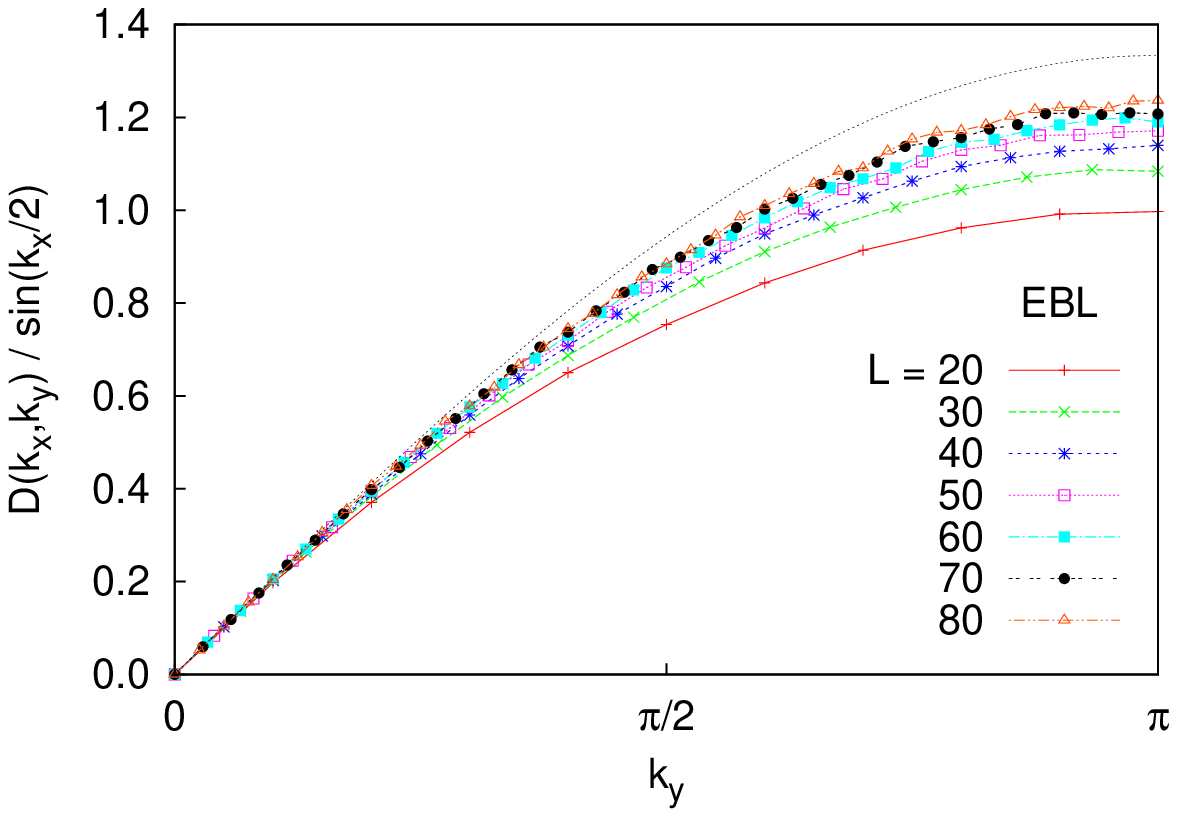}\\
  \includegraphics[width=\columnwidth]{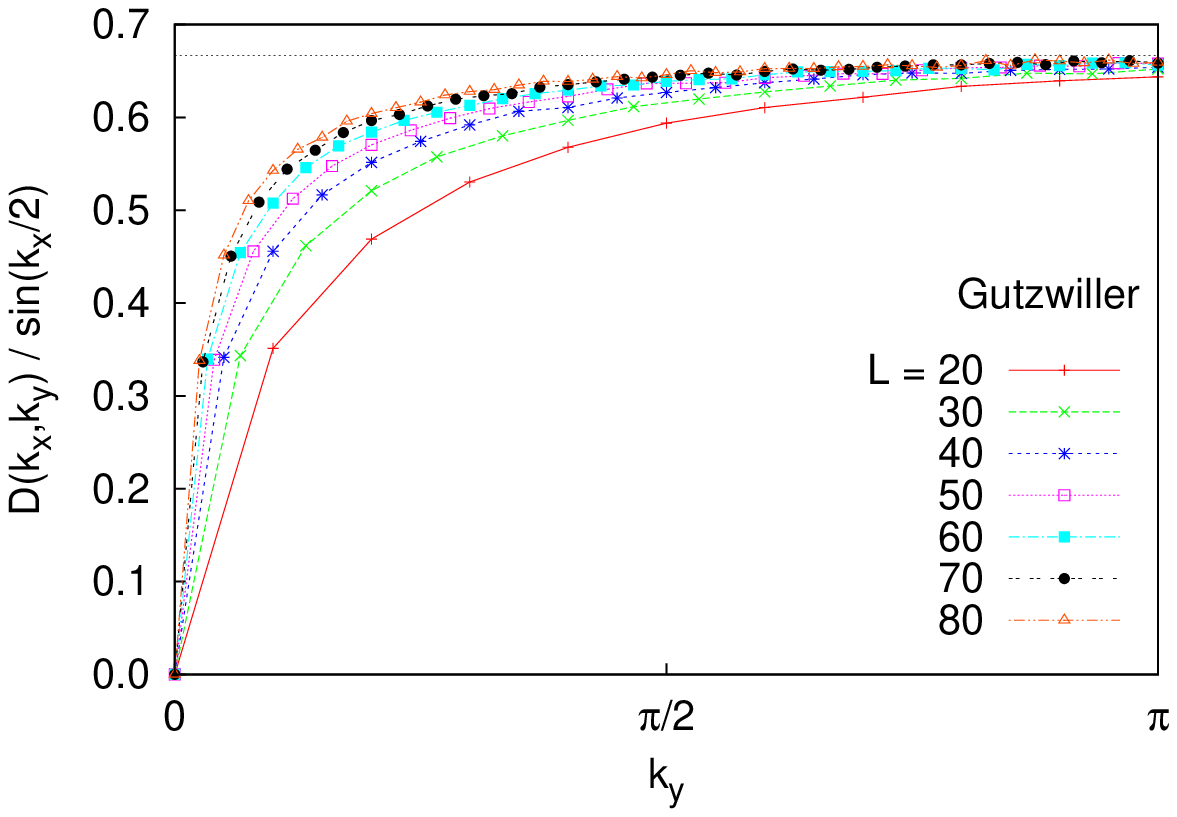}
  \caption{
Analysis of the VMC density structure factors at long wavelengths for the hard-core boson wave functions.  We plot $C(k_y) = D(k_x, k_y)/|\sin(k_x/2)|$ evaluated at the smallest $k_x=2\pi/L$ versus $k_y$ for system sizes from $L=20$ to $80$.  Top: The EBL result shows $C(k_y)$ approaching $|k_y|/W_\EBL$ for small $k_y$.  Bottom: The Gutzwiller result shows $C(k_y)$ approaching the constant $1/W_\Gutz$ for any fixed $k_y \neq 0$ upon increasing $L$.  The broken lines show the results obtained using harmonic approximations in the $L \to \infty$ limit.
}
  \label{fig:sf}
\end{figure}

We now present the results of our numerical study. In Fig.~\ref{fig:sf}, we analyze the density structure factor for each of the two wave functions by plotting
\begin{equation}
  C(k_y) = \left.\frac{D(k_x,k_y)}{\sin(k_x/2)} \right\vert_{k_x=2\pi/L},
\end{equation}
taken at the smallest $k_x=2\pi/L$.  This gives a finite-size measure of the slope of the density structure factor characterizing the V-shaped singularity in the small $k_x$ limit at fixed $k_y$, and we are further interested in the behavior of $C(k_y)$ for small $k_y$.  In the top panel, we obtain the limiting behavior for the EBL wave function $C_\EBL(k_y \to 0) \approx |k_y|/W_\EBL$, which agrees very well with the results derived using the harmonic approximation in Eq.~(\ref{eq:EBL_sf}) (illustrated as a broken line in the figure).

The bottom panel of Fig.~\ref{fig:sf} shows the corresponding analysis of the density structure factor for the Gutzwiller wave function. Here, $C_\Gutz(k_y)$ approaches constant $1/W_\Gutz$ (horizontal broken line) for any finite $k_y$ when lattice size $L \to \infty$, which again is in line with the result in the harmonic approximation in Eq.~(\ref{eq:sfGutz}). We also examine the ratio of the VMC Gutzwiller density structure factor to that in the harmonic approximation (not shown), and verify that the VMC data indeed converges toward the analytical trend in Eq.~(\ref{eq:sfGutz}) with increasing $L$. Thus, the density structure factor at long wavelengths clearly has a qualitatively different behavior for the EBL and Gutzwiller hard-core boson wave functions.

\begin{figure}
  \centering
  \includegraphics[width=\columnwidth]{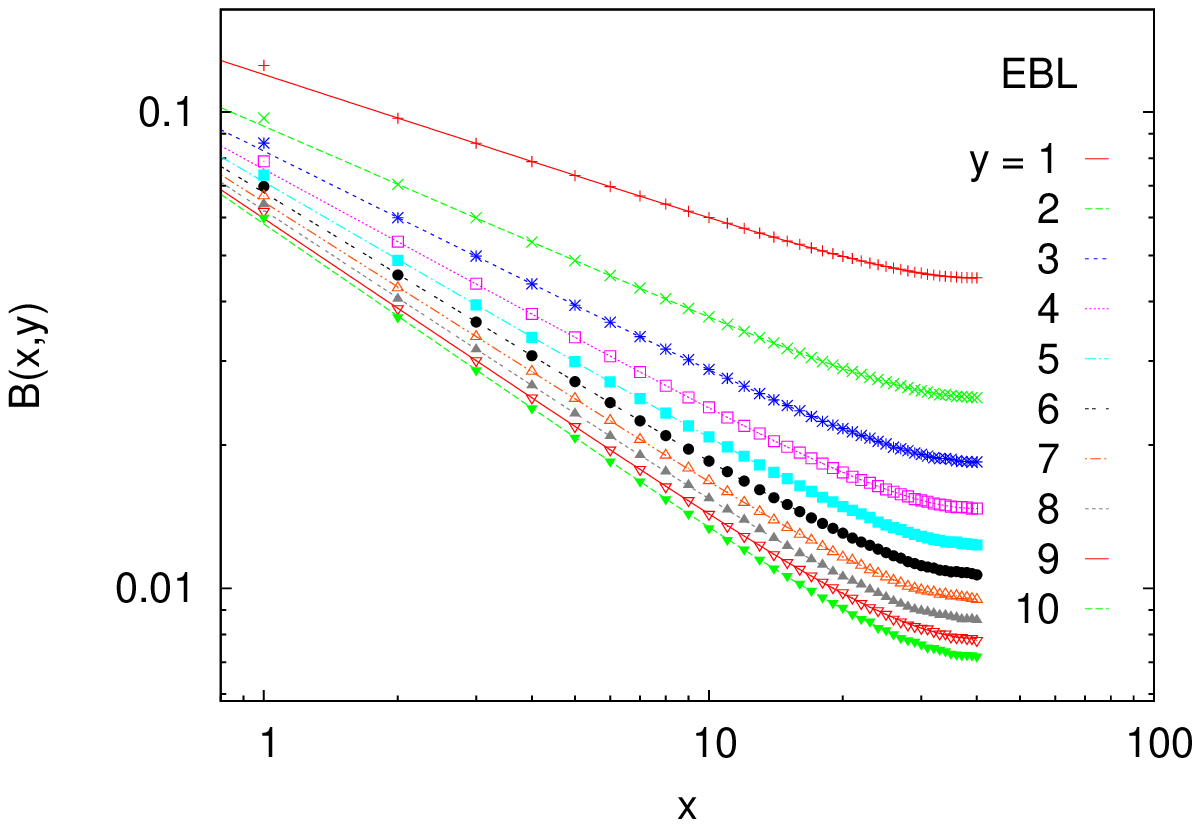}\\
  \includegraphics[width=\columnwidth]{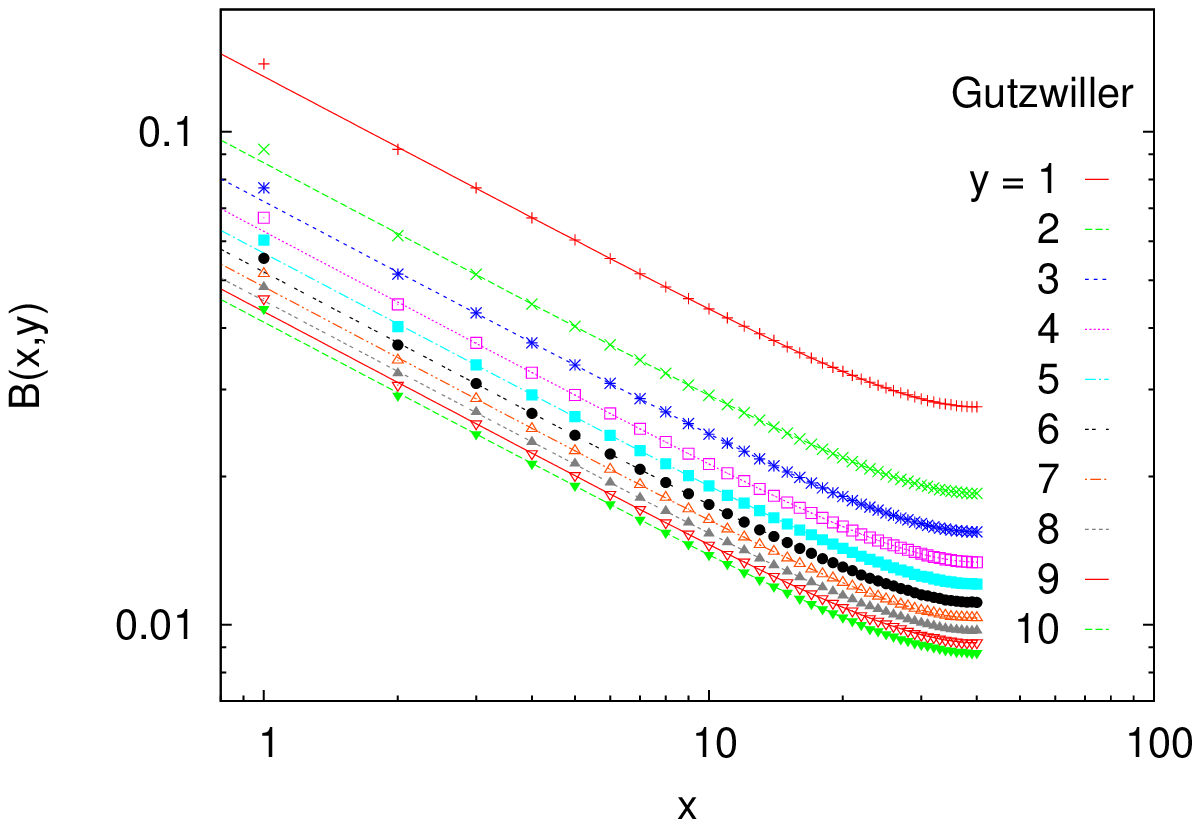}
  \caption{
Box correlator $\cB(x,y)$ versus $x$ for $y=1$ to 10, measured on a $80\times 80$ lattice. The data are fitted to the ansatz in Eq.~(\ref{eq:fit}), and the results show that the power law exponent $\eta(y)$ increases logarithmically with $y$ for the EBL wave function (top), while the exponent is essentially independent of $y$ for the Gutzwiller wave function (bottom).}
  \label{fig:boxCorr}
\end{figure}

Figure~\ref{fig:boxCorr} shows the box correlator $\cB(x,y)$ versus $x$ for $y=1$ to 10, measured on a $80\times 80$ lattice. The results for the EBL and Gutzwiller wave functions are given in the top and bottom panels respectively.  The data points are plotted together with the best-fit curves using the following ansatz
\begin{eqnarray}
  \cB(x,y) &=& A_y\left\vert\frac{L}{\pi}\sin\left(\frac{\pi x}{L}\right)\right\vert^{-\eta(y)},\label{eq:fit}\\
  &\approx& A_y \vert x\vert^{-\eta(y)}, ~~~~ x\ll L.
\end{eqnarray}
From the two plots shown in Fig.~\ref{fig:boxCorr}, it is clear that Eq.~(\ref{eq:fit}) provides very good fits for the data. For fixed $y$, the parameter $\eta(y)$ determines the exponent in the power-law relation $\cB(x,y)\sim|x|^{-\eta(y)}$.

For the EBL wave function, the lines fanning out in the top panel show that the fitting parameter $\eta(y)$ increases with $y$.  Anticipating a logarithmic relation from Eq.~(\ref{eq:log_approx}), we perform an additional data fit to the following
\begin{equation}
  \eta(y) = \gamma\log(y),
\end{equation}
and obtain $\gamma=0.148$. This value is very close to $W_\EBL/\pi^2$ from the harmonic approximation, and therefore suggests that the discreteness of the boson number and the hard-core repulsion in the wave function do not significantly alter the long wavelength properties of the resulting quantum state when the wave function is well inside the liquid regime.

For the Gutzwiller wave function, the lines running parallel to one another in the bottom panel in Fig.~\ref{fig:boxCorr} show that the exponent of the box correlator is essentially independent of $y$.  We obtain very good fits using $\eta(y)=0.477$, which is again very close to the corresponding Gaussian value $W_{\rm 1d}/\pi$. As before, the discreteness of the boson number and the hard-core repulsion do not modify the long wavelength results of the harmonic approximation.

We have thus shown that the EBL and Gutzwiller results obtained using the harmonic approximations in Sec.~\ref{sec:trial_wavefunction} remain valid for hard-core bosons here when the wave functions are well inside the liquid regimes.  But more importantly, the exact VMC density structure factor and box correlator show that the Gutzwiller projection leads to a quantum state that is qualitatively different from that of the EBL wave function.  We discuss further implications of this finding in the conclusion.

\section{Conclusions}\label{sec:conclusion}

In this work, we compared a Gutzwiller wave function with an EBL wave function, both motivated from the same parton-gauge action, and found that they realize quantum liquids with qualitatively different long-wavelength properties.  This shows that the Gutzwiller wave function, which does not include fluctuations of the spatial gauge field, has failed to capture the long-wavelength physics.  Similar approaches have often been used in the studies of quantum spin liquids and other strongly correlated systems, and in some cases, Gutzwiller wave functions with gapless partons possess competitive ground state energies.\cite{Motrunich2005, Yunoki06, HermeleRan2008, Heidarian2009, Clark2010, Motrunich2007, Sheng2008, Sheng2009, Block2010a, Block2010b}  We note that the gauge fluctuations in our case are more damped compared to the cases with generic parton Fermi surfaces\cite{LeeNagaosaWen2006, Holstein, Reizer, LeeNagaosa, Polchinski94, Altshuler94, YBKim94, SSLee2009, Metlitski2010, Mross2010} or Fermi points\cite{LeeNagaosaWen2006, Rantner2002, Hermele2004}, and hence are expected to be less important than in those cases, but still lead to qualitative effects as we have seen.  This therefore raises the possibility that Gutzwiller projection may generally fail to capture the correct ground state physics.

Let us also mention some extensions.  An interesting question in the same setting is to compare entanglement in the Gutzwiller and EBL wave functions,\cite{EE, Hastings2010} and examine the effect of including spatial gauge fluctuations as modeled by the latter.  A direct study both in the EBL field theory and in the hard core boson model realizations\cite{Tay2010a, Tay2010b} would be useful.

In this paper, we used bosonic partons, which we argued to be appropriate for the ring models with $K_{mn} > 0$; in this case the ground state wave function is positive and there is no sign problem.  On the other hand, for models with $K_{mn} < 0$, where in general there is a sign problem and the ground state wave function has non-trivial signs, it appears to be more appropriate to use fermionic partons, $b_{\br}^\dagger = d_{\br 1}^\dagger d_{\br 2}^\dagger$.\cite{Motrunich2007}  We can argue for this either from mean field energetics like in Eq.~(\ref{eq:ring_energy}), or from the connection between the corresponding $H_{\rm ring}$ and $H_{\rm U(1)}$.\cite{Motrunich2007}  This construction gives a so-called extremal DLBL state from Ref.~\onlinecite{Motrunich2007} where fermionic partons form flat Fermi surfaces in the mean field.  A naive bosonization treatment of the corresponding parton-gauge system leads to theory similar to our Eq.~(\ref{eq:parton_gauge}), and hence to an EBL-like long-wavelength description.  The Gutzwiller wave function $\Psi_b = \Psi_{d_1} \Psi_{d_2}$ is also similar to the one in the present study, but with specific sign structure from the product of the parton Slater determinants.  Inspired by the present work, we can attempt to crudely account for the gauge fluctuations by replacing the absolute value $|\Psi_{d_1} \Psi_{d_2}|$ by the Jastrow-EBL form while keeping the sign structure.  Interestingly, while the density correlations are not sensitive to the sign structure, the boson ring correlations are, and in the frustrated case they have faster power law decay (for the same density correlations), as can be seen already on the mean field level.

It would be interesting to examine other contexts with gapless parton-gauge systems where Gutzwiller-type wave functions have been used, such as gapless spin liquids\cite{LeeNagaosaWen2006, Motrunich2005, HermeleRan2008, Heidarian2009, Sheng2009, Block2010b, Clark2010} and more general Bose-metals\cite{Motrunich2007, Sheng2008, Block2010a}, and see if we can learn how to include gauge fluctuations in these cases, even if only on some crude level.

\acknowledgments
We would like to thank M. P. A. Fisher for many inspirations leading to this project, and to R. Kaul, M. Hastings, and A. Paramekanti for discussions.  The research is supported by the NSF through grant DMR-0907145 and the A.~P.~Sloan Foundation.

\bibliography{biblio}

\begin{thebibliography}{50}
\expandafter\ifx\csname natexlab\endcsname\relax\def\natexlab#1{#1}\fi
\expandafter\ifx\csname bibnamefont\endcsname\relax
  \def\bibnamefont#1{#1}\fi
\expandafter\ifx\csname bibfnamefont\endcsname\relax
  \def\bibfnamefont#1{#1}\fi
\expandafter\ifx\csname citenamefont\endcsname\relax
  \def\citenamefont#1{#1}\fi
\expandafter\ifx\csname url\endcsname\relax
  \def\url#1{\texttt{#1}}\fi
\expandafter\ifx\csname urlprefix\endcsname\relax\def\urlprefix{URL }\fi
\providecommand{\bibinfo}[2]{#2}
\providecommand{\eprint}[2][]{\url{#2}}

\bibitem[{\citenamefont{Lee et~al.}(2006)\citenamefont{Lee, Nagaosa, and
  Wen}}]{LeeNagaosaWen2006}
\bibinfo{author}{\bibfnamefont{P.~A.} \bibnamefont{Lee}},
  \bibinfo{author}{\bibfnamefont{N.}~\bibnamefont{Nagaosa}}, \bibnamefont{and}
  \bibinfo{author}{\bibfnamefont{X.-G.} \bibnamefont{Wen}},
  \bibinfo{journal}{Rev. Mod. Phys.} \textbf{\bibinfo{volume}{78}},
  \bibinfo{pages}{17} (\bibinfo{year}{2006}).

\bibitem[{\citenamefont{Balents}(2010)}]{Balents2010}
\bibinfo{author}{\bibfnamefont{L.}~\bibnamefont{Balents}},
  \bibinfo{journal}{Nature} \textbf{\bibinfo{volume}{464}},
  \bibinfo{pages}{199} (\bibinfo{year}{2010}),
  \urlprefix\url{http://dx.doi.org/10.1038/nature08917}.

\bibitem[{\citenamefont{Wen}(2002)}]{Wen2002}
\bibinfo{author}{\bibfnamefont{X.-G.} \bibnamefont{Wen}},
  \bibinfo{journal}{Phys. Rev. B} \textbf{\bibinfo{volume}{65}},
  \bibinfo{pages}{165113} (\bibinfo{year}{2002}).

\bibitem[{\citenamefont{Holstein et~al.}(1973)\citenamefont{Holstein, Norton,
  and Pincus}}]{Holstein}
\bibinfo{author}{\bibfnamefont{T.}~\bibnamefont{Holstein}},
  \bibinfo{author}{\bibfnamefont{R.~E.} \bibnamefont{Norton}},
  \bibnamefont{and} \bibinfo{author}{\bibfnamefont{P.}~\bibnamefont{Pincus}},
  \bibinfo{journal}{Phys. Rev. B} \textbf{\bibinfo{volume}{8}},
  \bibinfo{pages}{2649} (\bibinfo{year}{1973}).

\bibitem[{\citenamefont{Reizer}(1989)}]{Reizer}
\bibinfo{author}{\bibfnamefont{M.}~\bibnamefont{Reizer}},
  \bibinfo{journal}{Phys. Rev. B} \textbf{\bibinfo{volume}{40}},
  \bibinfo{pages}{11571} (\bibinfo{year}{1989}).

\bibitem[{\citenamefont{Lee and Nagaosa}(1992)}]{LeeNagaosa}
\bibinfo{author}{\bibfnamefont{P.~A.} \bibnamefont{Lee}} \bibnamefont{and}
  \bibinfo{author}{\bibfnamefont{N.}~\bibnamefont{Nagaosa}},
  \bibinfo{journal}{Phys. Rev. B} \textbf{\bibinfo{volume}{46}},
  \bibinfo{pages}{5621} (\bibinfo{year}{1992}).

\bibitem[{\citenamefont{Polchinski}(1994)}]{Polchinski94}
\bibinfo{author}{\bibfnamefont{J.}~\bibnamefont{Polchinski}},
  \bibinfo{journal}{Nucl. Phys. B} \textbf{\bibinfo{volume}{422}},
  \bibinfo{pages}{617} (\bibinfo{year}{1994}).

\bibitem[{\citenamefont{Altshuler et~al.}(1994)\citenamefont{Altshuler, Ioffe,
  and Millis}}]{Altshuler94}
\bibinfo{author}{\bibfnamefont{B.~L.} \bibnamefont{Altshuler}},
  \bibinfo{author}{\bibfnamefont{L.~B.} \bibnamefont{Ioffe}}, \bibnamefont{and}
  \bibinfo{author}{\bibfnamefont{A.~J.} \bibnamefont{Millis}},
  \bibinfo{journal}{Phys. Rev. B} \textbf{\bibinfo{volume}{50}},
  \bibinfo{pages}{14048} (\bibinfo{year}{1994}).

\bibitem[{\citenamefont{Kim et~al.}(1994)\citenamefont{Kim, Furusaki, Wen, and
  Lee}}]{YBKim94}
\bibinfo{author}{\bibfnamefont{Y.~B.} \bibnamefont{Kim}},
  \bibinfo{author}{\bibfnamefont{A.}~\bibnamefont{Furusaki}},
  \bibinfo{author}{\bibfnamefont{X.~G.} \bibnamefont{Wen}}, \bibnamefont{and}
  \bibinfo{author}{\bibfnamefont{P.~A.} \bibnamefont{Lee}},
  \bibinfo{journal}{Phys. Rev. B} \textbf{\bibinfo{volume}{50}},
  \bibinfo{pages}{17917} (\bibinfo{year}{1994}).

\bibitem[{\citenamefont{Lee}(2009)}]{SSLee2009}
\bibinfo{author}{\bibfnamefont{S.-S.} \bibnamefont{Lee}},
  \bibinfo{journal}{Phys. Rev. B} \textbf{\bibinfo{volume}{80}},
  \bibinfo{pages}{165102} (\bibinfo{year}{2009}).

\bibitem[{\citenamefont{Metlitski and Sachdev}(2010)}]{Metlitski2010}
\bibinfo{author}{\bibfnamefont{M.~A.} \bibnamefont{Metlitski}}
  \bibnamefont{and} \bibinfo{author}{\bibfnamefont{S.}~\bibnamefont{Sachdev}},
  \bibinfo{journal}{Phys. Rev. B} \textbf{\bibinfo{volume}{82}},
  \bibinfo{pages}{075127} (\bibinfo{year}{2010}).

\bibitem[{\citenamefont{Mross et~al.}(2010)\citenamefont{Mross, McGreevy, Liu,
  and Senthil}}]{Mross2010}
\bibinfo{author}{\bibfnamefont{D.~F.} \bibnamefont{Mross}},
  \bibinfo{author}{\bibfnamefont{J.}~\bibnamefont{McGreevy}},
  \bibinfo{author}{\bibfnamefont{H.}~\bibnamefont{Liu}}, \bibnamefont{and}
  \bibinfo{author}{\bibfnamefont{T.}~\bibnamefont{Senthil}},
  \bibinfo{journal}{Phys. Rev. B} \textbf{\bibinfo{volume}{82}},
  \bibinfo{pages}{045121} (\bibinfo{year}{2010}).

\bibitem[{\citenamefont{Rantner and Wen}(2002)}]{Rantner2002}
\bibinfo{author}{\bibfnamefont{W.}~\bibnamefont{Rantner}} \bibnamefont{and}
  \bibinfo{author}{\bibfnamefont{X.-G.} \bibnamefont{Wen}},
  \bibinfo{journal}{Phys. Rev. B} \textbf{\bibinfo{volume}{66}},
  \bibinfo{pages}{144501} (\bibinfo{year}{2002}).

\bibitem[{\citenamefont{Hermele et~al.}(2004)\citenamefont{Hermele, Senthil,
  Fisher, Lee, Nagaosa, and Wen}}]{Hermele2004}
\bibinfo{author}{\bibfnamefont{M.}~\bibnamefont{Hermele}},
  \bibinfo{author}{\bibfnamefont{T.}~\bibnamefont{Senthil}},
  \bibinfo{author}{\bibfnamefont{M.~P.~A.} \bibnamefont{Fisher}},
  \bibinfo{author}{\bibfnamefont{P.~A.} \bibnamefont{Lee}},
  \bibinfo{author}{\bibfnamefont{N.}~\bibnamefont{Nagaosa}}, \bibnamefont{and}
  \bibinfo{author}{\bibfnamefont{X.-G.} \bibnamefont{Wen}},
  \bibinfo{journal}{Phys. Rev. B} \textbf{\bibinfo{volume}{70}},
  \bibinfo{pages}{214437} (\bibinfo{year}{2004}).

\bibitem[{\citenamefont{Motrunich}(2005)}]{Motrunich2005}
\bibinfo{author}{\bibfnamefont{O.~I.} \bibnamefont{Motrunich}},
  \bibinfo{journal}{Phys. Rev. B} \textbf{\bibinfo{volume}{72}},
  \bibinfo{pages}{045105} (\bibinfo{year}{2005}).

\bibitem[{\citenamefont{Hermele et~al.}(2008)\citenamefont{Hermele, Ran, Lee,
  and Wen}}]{HermeleRan2008}
\bibinfo{author}{\bibfnamefont{M.}~\bibnamefont{Hermele}},
  \bibinfo{author}{\bibfnamefont{Y.}~\bibnamefont{Ran}},
  \bibinfo{author}{\bibfnamefont{P.~A.} \bibnamefont{Lee}}, \bibnamefont{and}
  \bibinfo{author}{\bibfnamefont{X.-G.} \bibnamefont{Wen}},
  \bibinfo{journal}{Phys. Rev. B} \textbf{\bibinfo{volume}{77}},
  \bibinfo{pages}{224413} (\bibinfo{year}{2008}).

\bibitem[{\citenamefont{Paramekanti et~al.}(2002)\citenamefont{Paramekanti,
  Balents, and Fisher}}]{Paramekanti2002}
\bibinfo{author}{\bibfnamefont{A.}~\bibnamefont{Paramekanti}},
  \bibinfo{author}{\bibfnamefont{L.}~\bibnamefont{Balents}}, \bibnamefont{and}
  \bibinfo{author}{\bibfnamefont{M.~P.~A.} \bibnamefont{Fisher}},
  \bibinfo{journal}{Phys. Rev. B} \textbf{\bibinfo{volume}{66}},
  \bibinfo{pages}{054526} (\bibinfo{year}{2002}).

\bibitem[{\citenamefont{Xu and Moore}(2005)}]{Xu2005}
\bibinfo{author}{\bibfnamefont{C.}~\bibnamefont{Xu}} \bibnamefont{and}
  \bibinfo{author}{\bibfnamefont{J.}~\bibnamefont{Moore}},
  \bibinfo{journal}{Nuclear Physics B} \textbf{\bibinfo{volume}{716}},
  \bibinfo{pages}{487 } (\bibinfo{year}{2005}).

\bibitem[{\citenamefont{Xu and Fisher}(2007)}]{Xu2007}
\bibinfo{author}{\bibfnamefont{C.}~\bibnamefont{Xu}} \bibnamefont{and}
  \bibinfo{author}{\bibfnamefont{M.~P.~A.} \bibnamefont{Fisher}},
  \bibinfo{journal}{Phys. Rev. B} \textbf{\bibinfo{volume}{75}},
  \bibinfo{pages}{104428} (\bibinfo{year}{2007}).

\bibitem[{\citenamefont{Tay and Motrunich}(2010)}]{Tay2010a}
\bibinfo{author}{\bibfnamefont{T.}~\bibnamefont{Tay}} \bibnamefont{and}
  \bibinfo{author}{\bibfnamefont{O.~I.} \bibnamefont{Motrunich}},
  \bibinfo{journal}{Phys. Rev. Lett.} \textbf{\bibinfo{volume}{105}},
  \bibinfo{pages}{187202} (\bibinfo{year}{2010}).

\bibitem[{\citenamefont{Tay and Motrunich}(arXiv:1011.0055)}]{Tay2010b}
\bibinfo{author}{\bibfnamefont{T.}~\bibnamefont{Tay}} \bibnamefont{and}
  \bibinfo{author}{\bibfnamefont{O.~I.} \bibnamefont{Motrunich}}
  (\bibinfo{year}{arXiv:1011.0055}).

\bibitem[{\citenamefont{Motrunich and Fisher}(2007)}]{Motrunich2007}
\bibinfo{author}{\bibfnamefont{O.~I.} \bibnamefont{Motrunich}}
  \bibnamefont{and} \bibinfo{author}{\bibfnamefont{M.~P.~A.}
  \bibnamefont{Fisher}}, \bibinfo{journal}{Phys. Rev. B}
  \textbf{\bibinfo{volume}{75}}, \bibinfo{pages}{235116}
  (\bibinfo{year}{2007}).

\bibitem[{\citenamefont{Nussinov and Fradkin}(2005)}]{Nussinov2005}
\bibinfo{author}{\bibfnamefont{Z.}~\bibnamefont{Nussinov}} \bibnamefont{and}
  \bibinfo{author}{\bibfnamefont{E.}~\bibnamefont{Fradkin}},
  \bibinfo{journal}{Phys. Rev. B} \textbf{\bibinfo{volume}{71}},
  \bibinfo{pages}{195120} (\bibinfo{year}{2005}).

\bibitem[{\citenamefont{Batista and Nussinov}(2005)}]{Batista2005}
\bibinfo{author}{\bibfnamefont{C.~D.} \bibnamefont{Batista}} \bibnamefont{and}
  \bibinfo{author}{\bibfnamefont{Z.}~\bibnamefont{Nussinov}},
  \bibinfo{journal}{Phys. Rev. B} \textbf{\bibinfo{volume}{72}},
  \bibinfo{pages}{045137} (\bibinfo{year}{2005}).

\bibitem[{\citenamefont{Nussinov et~al.}(2006)\citenamefont{Nussinov, Batista,
  and Fradkin}}]{Nussinov2006}
\bibinfo{author}{\bibfnamefont{Z.}~\bibnamefont{Nussinov}},
  \bibinfo{author}{\bibfnamefont{C.~D.} \bibnamefont{Batista}},
  \bibnamefont{and} \bibinfo{author}{\bibfnamefont{E.}~\bibnamefont{Fradkin}},
  \bibinfo{journal}{Int. Journ. Mod. Phys. B} \textbf{\bibinfo{volume}{20}},
  \bibinfo{pages}{5239} (\bibinfo{year}{2006}).

\bibitem[{\citenamefont{Ceperley et~al.}(1977)\citenamefont{Ceperley, Chester,
  and Kalos}}]{Ceperley77}
\bibinfo{author}{\bibfnamefont{D.~M.} \bibnamefont{Ceperley}},
  \bibinfo{author}{\bibfnamefont{G.~V.} \bibnamefont{Chester}},
  \bibnamefont{and} \bibinfo{author}{\bibfnamefont{M.~H.} \bibnamefont{Kalos}},
  \bibinfo{journal}{Phys. Rev. B} \textbf{\bibinfo{volume}{16}},
  \bibinfo{pages}{3081} (\bibinfo{year}{1977}).

\bibitem[{\citenamefont{Gros}(1989)}]{Gros89}
\bibinfo{author}{\bibfnamefont{C.}~\bibnamefont{Gros}},
  \bibinfo{journal}{Annals Phys. (NY)} \textbf{\bibinfo{volume}{189}},
  \bibinfo{pages}{53} (\bibinfo{year}{1989}).

\bibitem[{\citenamefont{Haldane}(1981)}]{Haldane1981}
\bibinfo{author}{\bibfnamefont{F.~D.~M.} \bibnamefont{Haldane}},
  \bibinfo{journal}{Phys. Rev. Lett.} \textbf{\bibinfo{volume}{47}},
  \bibinfo{pages}{1840} (\bibinfo{year}{1981}).

\bibitem[{\citenamefont{Fisher and Lee}(1989)}]{Fisher1989}
\bibinfo{author}{\bibfnamefont{M.~P.~A.} \bibnamefont{Fisher}}
  \bibnamefont{and} \bibinfo{author}{\bibfnamefont{D.~H.} \bibnamefont{Lee}},
  \bibinfo{journal}{Phys. Rev. B} \textbf{\bibinfo{volume}{39}},
  \bibinfo{pages}{2756} (\bibinfo{year}{1989}).

\bibitem[{\citenamefont{Fisher}(arXiv:Cond-mat/9806164v2)}]{Fisher1998}
\bibinfo{author}{\bibfnamefont{M.~P.~A.} \bibnamefont{Fisher}}
  (\bibinfo{year}{arXiv:Cond-mat/9806164v2}).

\bibitem[{\citenamefont{Giamarchi}(2004)}]{Giamarchi}
\bibinfo{author}{\bibfnamefont{T.}~\bibnamefont{Giamarchi}},
  \emph{\bibinfo{title}{Quantum Physics in One Dimension}}
  (\bibinfo{publisher}{Clarendon Press, Oxford}, \bibinfo{year}{2004}).

\bibitem[{\citenamefont{Emery et~al.}(2000)\citenamefont{Emery, Fradkin,
  Kivelson, and Lubensky}}]{Emery2000}
\bibinfo{author}{\bibfnamefont{V.~J.} \bibnamefont{Emery}},
  \bibinfo{author}{\bibfnamefont{E.}~\bibnamefont{Fradkin}},
  \bibinfo{author}{\bibfnamefont{S.~A.} \bibnamefont{Kivelson}},
  \bibnamefont{and} \bibinfo{author}{\bibfnamefont{T.~C.}
  \bibnamefont{Lubensky}}, \bibinfo{journal}{Phys. Rev. Lett.}
  \textbf{\bibinfo{volume}{85}}, \bibinfo{pages}{2160} (\bibinfo{year}{2000}).

\bibitem[{\citenamefont{Vishwanath and Carpentier}(2001)}]{Vishwanath2001}
\bibinfo{author}{\bibfnamefont{A.}~\bibnamefont{Vishwanath}} \bibnamefont{and}
  \bibinfo{author}{\bibfnamefont{D.}~\bibnamefont{Carpentier}},
  \bibinfo{journal}{Phys. Rev. Lett.} \textbf{\bibinfo{volume}{86}},
  \bibinfo{pages}{676} (\bibinfo{year}{2001}).

\bibitem[{\citenamefont{Mukhopadhyay et~al.}(2001)\citenamefont{Mukhopadhyay,
  Kane, and Lubensky}}]{Ranjan2001long}
\bibinfo{author}{\bibfnamefont{R.}~\bibnamefont{Mukhopadhyay}},
  \bibinfo{author}{\bibfnamefont{C.~L.} \bibnamefont{Kane}}, \bibnamefont{and}
  \bibinfo{author}{\bibfnamefont{T.~C.} \bibnamefont{Lubensky}},
  \bibinfo{journal}{Phys. Rev. B} \textbf{\bibinfo{volume}{64}},
  \bibinfo{pages}{045120} (\bibinfo{year}{2001}).

\bibitem[{\citenamefont{Bijl}(1940)}]{Bijl1940}
\bibinfo{author}{\bibfnamefont{A.}~\bibnamefont{Bijl}},
  \bibinfo{journal}{Physica} \textbf{\bibinfo{volume}{7}}, \bibinfo{pages}{869
  } (\bibinfo{year}{1940}).

\bibitem[{\citenamefont{Jastrow}(1955)}]{Jastrow1955}
\bibinfo{author}{\bibfnamefont{R.}~\bibnamefont{Jastrow}},
  \bibinfo{journal}{Phys. Rev.} \textbf{\bibinfo{volume}{98}},
  \bibinfo{pages}{1479} (\bibinfo{year}{1955}).

\bibitem[{\citenamefont{Reatto and Chester}(1967)}]{Reatto1967}
\bibinfo{author}{\bibfnamefont{L.}~\bibnamefont{Reatto}} \bibnamefont{and}
  \bibinfo{author}{\bibfnamefont{G.~V.} \bibnamefont{Chester}},
  \bibinfo{journal}{Phys. Rev.} \textbf{\bibinfo{volume}{155}},
  \bibinfo{pages}{88} (\bibinfo{year}{1967}).

\bibitem[{\citenamefont{Hellberg and Mele}(1991)}]{Hellberg1991}
\bibinfo{author}{\bibfnamefont{C.~S.} \bibnamefont{Hellberg}} \bibnamefont{and}
  \bibinfo{author}{\bibfnamefont{E.~J.} \bibnamefont{Mele}},
  \bibinfo{journal}{Phys. Rev. Lett.} \textbf{\bibinfo{volume}{67}},
  \bibinfo{pages}{2080} (\bibinfo{year}{1991}).

\bibitem[{\citenamefont{Capello et~al.}(2005)\citenamefont{Capello, Becca,
  Yunoki, Fabrizio, and Sorella}}]{Capello2005}
\bibinfo{author}{\bibfnamefont{M.}~\bibnamefont{Capello}},
  \bibinfo{author}{\bibfnamefont{F.}~\bibnamefont{Becca}},
  \bibinfo{author}{\bibfnamefont{S.}~\bibnamefont{Yunoki}},
  \bibinfo{author}{\bibfnamefont{M.}~\bibnamefont{Fabrizio}}, \bibnamefont{and}
  \bibinfo{author}{\bibfnamefont{S.}~\bibnamefont{Sorella}},
  \bibinfo{journal}{Phys. Rev. B} \textbf{\bibinfo{volume}{72}},
  \bibinfo{pages}{085121} (\bibinfo{year}{2005}).

\bibitem[{Sin()}]{SingleBosonHopping}
\bibinfo{note}{Although the row and column boson numbers are not strictly
  enforced in the harmonic approximations for both wave functions, the presence
  of similar ``cross'' feature in the density structure factors in
  Eqs.~(\ref{eq:EBL_sf}) and (\ref{eq:sfGutz}) suggests that this still holds
  approximately. To show this more explicitly, we calculate the single boson
  propagator for both wave functions and obtain the form $\la b^\dagger_{\bf 0}
  b_\br \ra \sim \delta_{\br, {\bf 0}}$. Interestingly, this shows that single
  boson hopping is suppressed even in the harmonic approximations.}

\bibitem[{Wcr()}]{Wcritical}
\bibinfo{note}{Our estimates are $W_{\rm EBL}^{\rm crit} \approx 4.4$ and
  $W_{\rm 1d}^{\rm crit} \approx 3.4$ for developing $(\pi,\pi)$ CDW order at
  half-filling in the EBL and Gutzwiller wave functions respectively.}

\bibitem[{\citenamefont{Yunoki and Sorella}(2006)}]{Yunoki06}
\bibinfo{author}{\bibfnamefont{S.}~\bibnamefont{Yunoki}} \bibnamefont{and}
  \bibinfo{author}{\bibfnamefont{S.}~\bibnamefont{Sorella}},
  \bibinfo{journal}{Phys. Rev. B} \textbf{\bibinfo{volume}{74}},
  \bibinfo{pages}{014408} (\bibinfo{year}{2006}).

\bibitem[{\citenamefont{Heidarian et~al.}(2009)\citenamefont{Heidarian,
  Sorella, and Becca}}]{Heidarian2009}
\bibinfo{author}{\bibfnamefont{D.}~\bibnamefont{Heidarian}},
  \bibinfo{author}{\bibfnamefont{S.}~\bibnamefont{Sorella}}, \bibnamefont{and}
  \bibinfo{author}{\bibfnamefont{F.}~\bibnamefont{Becca}},
  \bibinfo{journal}{Phys. Rev. B} \textbf{\bibinfo{volume}{80}},
  \bibinfo{pages}{012404} (\bibinfo{year}{2009}).

\bibitem[{\citenamefont{{Clark} et~al.}(arXiv:1010.3011)\citenamefont{{Clark},
  {Abanin}, and {Sondhi}}}]{Clark2010}
\bibinfo{author}{\bibfnamefont{B.~K.} \bibnamefont{{Clark}}},
  \bibinfo{author}{\bibfnamefont{D.~A.} \bibnamefont{{Abanin}}},
  \bibnamefont{and} \bibinfo{author}{\bibfnamefont{S.~L.}
  \bibnamefont{{Sondhi}}} (\bibinfo{year}{arXiv:1010.3011}).

\bibitem[{\citenamefont{Sheng et~al.}(2008)\citenamefont{Sheng, Motrunich,
  Trebst, Gull, and Fisher}}]{Sheng2008}
\bibinfo{author}{\bibfnamefont{D.~N.} \bibnamefont{Sheng}},
  \bibinfo{author}{\bibfnamefont{O.~I.} \bibnamefont{Motrunich}},
  \bibinfo{author}{\bibfnamefont{S.}~\bibnamefont{Trebst}},
  \bibinfo{author}{\bibfnamefont{E.}~\bibnamefont{Gull}}, \bibnamefont{and}
  \bibinfo{author}{\bibfnamefont{M.~P.~A.} \bibnamefont{Fisher}},
  \bibinfo{journal}{Phys. Rev. B} \textbf{\bibinfo{volume}{78}},
  \bibinfo{pages}{054520} (\bibinfo{year}{2008}).

\bibitem[{\citenamefont{Sheng et~al.}(2009)\citenamefont{Sheng, Motrunich, and
  Fisher}}]{Sheng2009}
\bibinfo{author}{\bibfnamefont{D.~N.} \bibnamefont{Sheng}},
  \bibinfo{author}{\bibfnamefont{O.~I.} \bibnamefont{Motrunich}},
  \bibnamefont{and} \bibinfo{author}{\bibfnamefont{M.~P.~A.}
  \bibnamefont{Fisher}}, \bibinfo{journal}{Phys. Rev. B}
  \textbf{\bibinfo{volume}{79}}, \bibinfo{pages}{205112}
  (\bibinfo{year}{2009}).

\bibitem[{\citenamefont{{Block} et~al.}(arXiv:1008.4105)\citenamefont{{Block},
  {Mishmash}, {Kaul}, {Sheng}, {Motrunich}, and {Fisher}}}]{Block2010a}
\bibinfo{author}{\bibfnamefont{M.~S.} \bibnamefont{{Block}}},
  \bibinfo{author}{\bibfnamefont{R.~V.} \bibnamefont{{Mishmash}}},
  \bibinfo{author}{\bibfnamefont{R.~K.} \bibnamefont{{Kaul}}},
  \bibinfo{author}{\bibfnamefont{D.~N.} \bibnamefont{{Sheng}}},
  \bibinfo{author}{\bibfnamefont{O.~I.} \bibnamefont{{Motrunich}}},
  \bibnamefont{and} \bibinfo{author}{\bibfnamefont{M.~P.~A.}
  \bibnamefont{{Fisher}}} (\bibinfo{year}{arXiv:1008.4105}).

\bibitem[{\citenamefont{{Block} et~al.}(arXiv:1009.1179)\citenamefont{{Block},
  {Sheng}, {Motrunich}, and {Fisher}}}]{Block2010b}
\bibinfo{author}{\bibfnamefont{M.~S.} \bibnamefont{{Block}}},
  \bibinfo{author}{\bibfnamefont{D.~N.} \bibnamefont{{Sheng}}},
  \bibinfo{author}{\bibfnamefont{O.~I.} \bibnamefont{{Motrunich}}},
  \bibnamefont{and} \bibinfo{author}{\bibfnamefont{M.~P.~A.}
  \bibnamefont{{Fisher}}} (\bibinfo{year}{arXiv:1009.1179}).

\bibitem[{\citenamefont{Calabrese et~al.}(2009)\citenamefont{Calabrese, Cardy,
  and Doyon}}]{EE}
\bibinfo{author}{\bibfnamefont{P.}~\bibnamefont{Calabrese}},
  \bibinfo{author}{\bibfnamefont{J.}~\bibnamefont{Cardy}}, \bibnamefont{and}
  \bibinfo{author}{\bibfnamefont{B.}~\bibnamefont{Doyon}},
  \bibinfo{journal}{Journal of Physics A: Mathematical and Theoretical}
  \textbf{\bibinfo{volume}{42}}, \bibinfo{pages}{500301}
  (\bibinfo{year}{2009}).

\bibitem[{\citenamefont{Hastings et~al.}(2010)\citenamefont{Hastings,
  Gonz\'alez, Kallin, and Melko}}]{Hastings2010}
\bibinfo{author}{\bibfnamefont{M.~B.} \bibnamefont{Hastings}},
  \bibinfo{author}{\bibfnamefont{I.}~\bibnamefont{Gonz\'alez}},
  \bibinfo{author}{\bibfnamefont{A.~B.} \bibnamefont{Kallin}},
  \bibnamefont{and} \bibinfo{author}{\bibfnamefont{R.~G.} \bibnamefont{Melko}},
  \bibinfo{journal}{Phys. Rev. Lett.} \textbf{\bibinfo{volume}{104}},
  \bibinfo{pages}{157201} (\bibinfo{year}{2010}).

\end{thebibliography}

\end{document}